\pdfoutput=1
\documentclass[prb,showpacs,twocolumn]{revtex4}
\usepackage{graphicx,amssymb,amsmath}
\begin{document}
\title{Measurement of the Transmission Phase through a Quantum Dot Embedded in One Arm
of an Electronic Mach-Zehnder Interferometer}
\author{L. V. Litvin }
\author{A. Helzel }
\author{H.-P. Tranitz }
\author{W. Wegscheider }
\author{C. Strunk}
\affiliation{ \mbox{Institut f\"{u}r experimentelle und angewandte
Physik, Universit\"{a}t  Regensburg, D-93040 Regensburg, Germany}}
\date{\today}
\begin{abstract}
We investigate an electronic Mach-Zehnder interferometer with high
visibility in the quantum Hall regime. The superposition of the
electrostatic potentials from a quantum point contact (QPC) and the
residual disorder potential from doping impurities frequently
results in the formation of inadvertent quantum dots (QD) in one arm
of the interferometer. This gives rise to resonances in the QPC
transmission characteristics. While crossing the QD resonance in
energy, the interferometer gains a phase shift of $\pi$ in
the interference pattern.
\end{abstract}
\pacs{73.23.Ad, 73.63.Nm}
%
%
\maketitle

\section{Introduction}

The conductance $G$ of an Aharonov-Bohm (AB) ring oscillates with
magnetic field $B$ with a period $\Delta B$ determined by the area $A$
between two interfering paths $\Delta B=\Phi_{0}/A$, where
$\Phi_{0}=h/e$ is the flux quantum. The combination of the AB ring
with a quantum dot (QD) in one of its arms gives the opportunity to
measure the phase of transmission amplitude through the QD.
\cite{yacoby} The dot, tuned to a transmission resonance, sustains
coherent transport all over the width of the resonance peak. So far this
was realized in different groups by recording AB oscillations as a function of the
magnetic field and tracing their phase with respect to the energy of
the resonant state, which was controlled by a plunger gate.
\cite{shuster,buks,kobayashi}
Fano resonances were observed in Ref.~\onlinecite{kobayashi}, while other works \cite{shuster,buks} show the phase evolution of AB oscillations while scanning through each Coulomb peak in energy changing the plunger gate. Here a slip by $\pi$ was seen, what can be explained in a single particle picture. For the observed $\pi$-jump between two Coulomb peaks\cite{shuster} a theory involving many particles is needed.
Besides the phase of the transmitted current also the phase of the reflected current was probed \cite{buks}, which showed similar results. In order to determine the energy-dependence of the
phase, it was necessary to record many $G(B)$-traces differing in
plunger gate, each resulting, in a single point in the phase
evolution. The question arises, whether one could directly measure
the phase in an interferometer, when the QD crosses a resonance. This
has two attractive advantages. First the measurement process speeds
up, because information about phase is acquired in a single sweep of
the plunger gate. Second, with such a fast measurement process, one
could think about detection of the charge state for the QD by
measuring its transmission (reflection) phase. The latter can be
important for building charge qubits based on double quantum dot
system. However in conventional AB interferometers, the small
interference contrast (typically 10\%) and signal noise makes this
task difficult.\cite{kobayashi} In this work we report on measurements of QD
transmission phase with an electronic Mach-Zehnder
interferometer.\cite{ji} The electronic Mach-Zehnder interferometer
employs edge channels of a two dimensional electron system in the
quantum Hall regime and quantum point contacts as beam splitters.
The interference contrast can be very large, up to 80~\%, at
temperatures near 20 mK.\cite{neder}

%
\begin{figure}
\includegraphics[width=85mm]{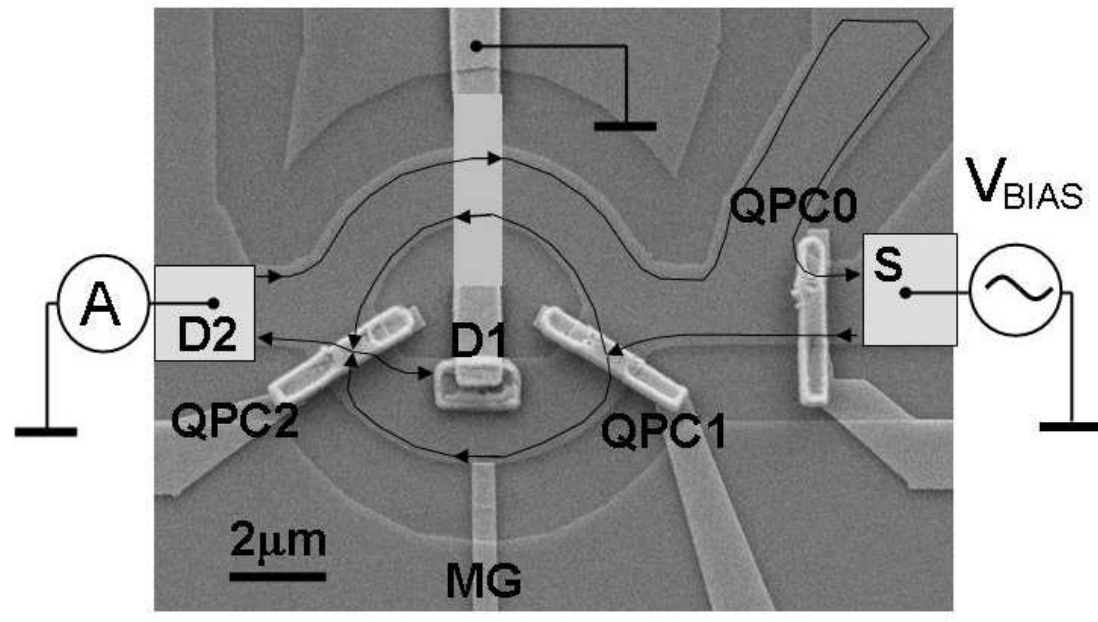}
\caption{SEM image of Mach-Zehnder interferometer with the scheme of
paths for nonequilibrium current. The transmission of QPC1 and QPC2
is set to 0.5. QPC0 transmits the outer edge channel and reflects
the inner one in case the filling factor being more than 1.
The modulation gate (MG) is used to shift the phase.}
\end{figure}

\section{Experimental Details}
The interferometer (see Fig.~1) was fabricated on the basis of a
modulation doped GaAs/Ga$_{x}$Al$_{1-x}$As heterostructure
containing a two-dimensional electron gas (2DEG) 90~nm below the
surface. At~4~K, the unpatterned 2DEG density and mobility were
$n$=2.0$\times$10$^{15}$~m$^{-2}$ and $\mu$=206~m$^{2}/$(Vs),
respectively. The details of fabrication procedure can be found in
Ref.~\onlinecite{litvin1}. Each arm of the MZI was approximately 9~$\mu$m
long and the gap between the tips of quantum point contacts was 400~nm.
This interferometer showed a maximum visibility of 56\% and an area  of
25~$\mu m^{2}$ between interfering paths, found from the period of
Aharonov-Bohm oscillations. A standard lock-in technique
($f\sim\;$300~Hz) with 1~$\mu$V excitation at terminal S and
detection at terminal D2 was employed (see Fig.~1). All measurements
were performed at a temperature below 50~mK.

\begin{figure}
\includegraphics[width=85mm]{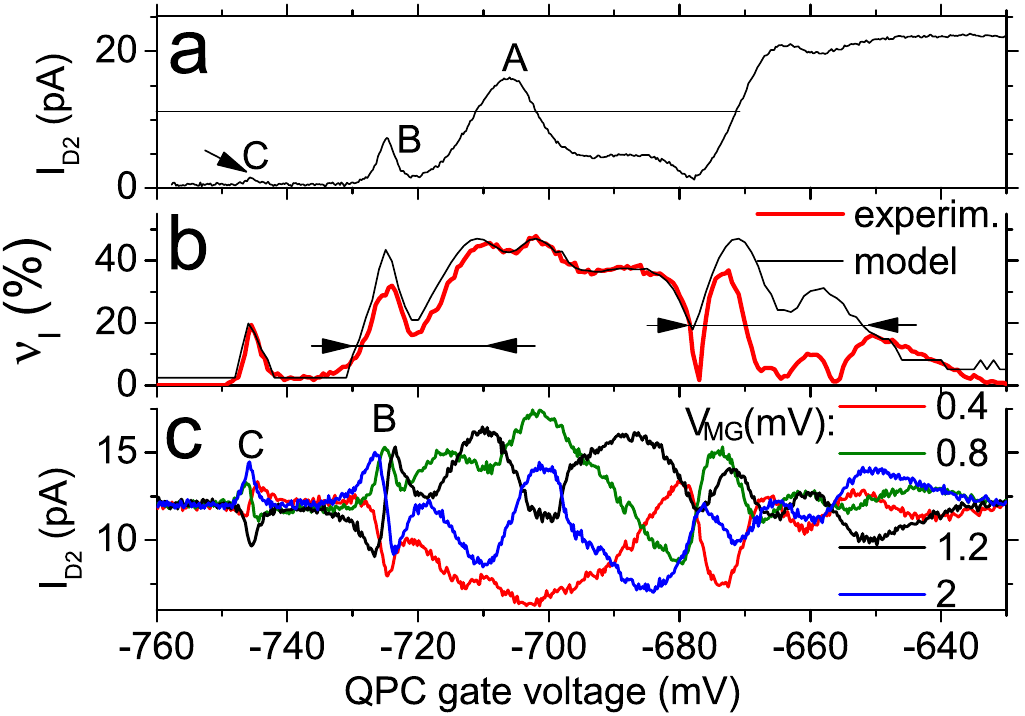}
\caption{(a) Transmission characteristic of QPC2 as a
function of the QPC's gate voltages for at $B$=4.6T. The horizontal line denotes half transmission of the QPC which should in a single particle picture correspond to the highest visibility. (b) Visibility
of Aharonov-Bohm oscillations, experimentally measured (thick red
line) and calculated from transmission of QPC2 (thin black line),
versus QPC2 gate voltage ($T_{QPC1}=0.5$, see Eq.~\ref{T}). The
regions of discrepancy are marked by arrows. (c) Interferometer
current for different modulation gate voltages, e.i. different AB-phases ($T_{QPC1}$=0.5). }
\end{figure}

\section{Characterization of the Quantum Point Contacts}
In many experiments, the transmission characteristics of the QPCs,
in high perpendicular magnetic field exhibit resonances superimposed
on the conductance steps [Fig.~2(a), see also
\cite{neder1,roulleau,litvin2}]. As we will show below, our data
suggest that this originates from the Coulomb blockade of a
quantum dot formed inadvertently by the disorder potential in the
vicinity of the QPCs. The interference contrast was highest for the
outer edge channel. To record the characteristics in Fig.~2(a), (i)
the magnetic field was set to 4.6~T (filling factor 1.6), (ii) QPC0
was adjusted to transmit only the outermost channel, and (iii) the
gate voltage for QPC2 was swept  to negative voltages, while
keeping QPC1 open [see also Fig.~3(a)].
A sequence of peaks (marked by letters A, B, C in Fig.~2(a) in the
gate characteristics of QPC2 appears at transmissions less than 1.
It is of interest to check if any peak structure can be found in the dependence on a magnetic field. The magnetic
field dependence of current transmitted through QPC2 is plotted in Fig.~3(c). While sweeping
the magnetic field, the data was recorded after adjusting QPC0 to
reflect the half-filled upper Landau level, opening QPC1, and tuning
the gate voltage of QPC2 at the maximum of peak "A"
($T_{QPC2}=0.75$) ($B$=4.6T).
The
average current in Fig.~3(c) corresponding to
$T_{QPC2}=0.4$ at $B$=4.8T increases to the left from that point,
reaching $T_{QPC2}=1$ at $B$=3.8T, and decreases to the right
approaching value $T_{QPC2}\approx0.1$. This occurs because of the
change of energy for the lowest Landau subband. In other words the
potential barrier adjusted at $B$=4.8T decreases and disappears to
the left from this point and grows to the right as function of $B$.
We find that this current has oscillatory components, which are not
expected for a single barrier but could easily appear in a device with
two and more barriers. Fourier analysis reveals two frequencies in
$B$, corresponding to periods of 0.27 and 0.18~T, and their higher
harmonics. When interpreted as Aharonov-Bohm oscillations ($\Delta
B\cdot A=e/h $) these two periods correspond to areas enclosed by
circumference with diameter $d=2\sqrt{A/\pi}$ of 140 and 170~nm.
This is well compatible with the lithographic gap between QPC tips
(400nm) and with the spatial variation of the disorder potential
near 100 nm measured, e.g., in Ref.~\onlinecite{steele} for 2DEG
structure with parameters similar to ours. Aharonov-Bohm
oscillations induced by potential fluctuations in single QPC in high
magnetic field was reported before.~\cite{loosdr} In addition, we
found that the resonances in Fig.~2(a) are also dependent on cooling
cycle, i.e. the shape of the resonances is unique for each cool
down. This indicates that charging of donor atoms in the doping
layer plays a role in the resonance formation.

\begin{figure}
\includegraphics[width=85mm]{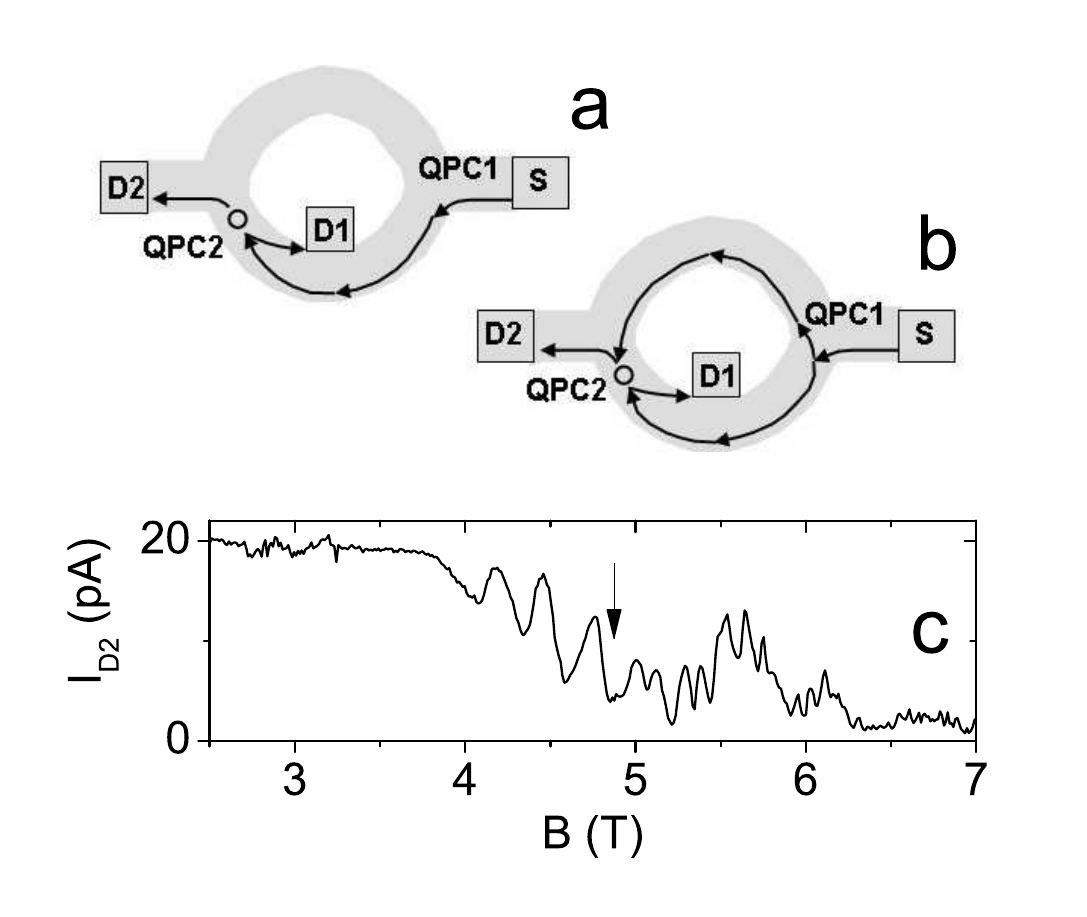}
\caption{(a) Scheme for probing transmission characteristic of QPC2
with QPC1 opened, and that (b) for interference experiment with both
QPCs partially transmitted. The circle at the QPC2 symbolizes the
quasi-bound state of quantum dot. (c) Oscillatory current of
outermost edge channel as function of magnetic field when
transmission of QPC2 is less than 1 (full line). An expected full
current for an outer channel is shown by crosses. Arrow marks the
point where new harmonic sets in.}
\end{figure}

\section{Phase shifts by the Quantum Dot}
Next, we discuss the response of the MZI interference near the
resonances in the transmission of QPC2. To investigate this, first,
the QPC1 must be set to half transmission, generating two
interfering paths as shown in Fig.~3(b). Second, the interference
signal must be measured as a function of the QPC2 gate voltage. Here
two possibilities arise depending on the regime for a modulation gate
voltage $V_{MG}$ which is normally used to observe interference by
shifting the phase of one arm with respect to other. These are i)
sweeping $V_{MG}$ simultaneously with the QPC2 gate voltage; and ii)
keeping $V_{MG}$ constant. The former allows to determine the
interference contrast as function of $V_{QPC2}$ (or $T_{QPC2}$) and
was demonstrated before [Ref.~\onlinecite{litvin2}, Fig.~1(c);
Fig.~2(a)]. In contrast, the latter is sensitive to any phase gain
during a change of the QPC2 gate voltage. We explore both of these
opportunities starting from the measurement of the span for AB
oscillation vs. $V_{QPC2}$. The relative amplitude of oscillations
is called visibility $\nu_{I}$,
$\nu_{I}~=~(I_{max}-I_{min})/(I_{max}+I_{min})$, and is plotted in
Fig.~2(b). Here one sees that the visibility $\nu_{I}$ peaks at the
transmission resonances "A", "B" and "C". Qualitatively, this is
explained by the fact that the visibility has a maximum when both
arms carry equal current, i.e. at $T_{QPC}=1/2$ (a quantitative
analysis follows below). Therefore the closer  $T_\text{QPC}$ is to
$1/2$, the higher is the visibility. The phase information does not
show itself in this experiment, but it does when the modulation gate
voltage $V_{MG}$ is kept constant [Fig.~2(c)]. We focus on the
resonances "B" and "C" where an abrupt change of the current is
observed. Some traces show at "B" and "C" a clear alternation between a peak and a dip structure. That means if the AB oscillations are adjusted to a maximum with MG, it changes abruptly to a minimum, when passing the resonance. This implies a phase change of $\pi$ at the resonance.

Now we address quantitatively the results in Fig.~2 in the framework
of a model assuming noninteracting particles. This model predicts the
MZI transmission coefficient \cite{ji}
\begin{equation} T_{SD2}=\vert t_{1}r_{2}\vert ^{2}+\vert r_{1}t_{2}\vert^{2}-2\vert
t_{1}t_{2}r_{1}r_{2}\vert \cos\Delta\varphi\;,\end{equation} here
$t_{i}, r_{i}$ are transmission and reflection amplitudes of QPCs
and $\Delta\varphi$ is the phase difference between the interferometer arms.
From this formula one can easily find the expression for the
visibility as function of one of the QPC transmission, namely,
\begin{equation}\label{T}
\nu_{I}=z\cdot2\sqrt{T_{QPC2}(1-T_{QPC2})},\end{equation}

where the
factor $z<1$
 accounts for the decoherence at finite temperature.
 Using this expression and the measured transmission
values for $T_{QPC2}$ in Fig.~2(a), we calculate the dependence of
visibility $\nu_{I}(T_{QPC2})$ and compare it with the experimental one.
The black, thin line in Fig.~2(b) shows the result, which agrees in
general well with the experiment. There are two regions of deviation
from the simple model of Eq.~\ref{T}, marked by arrows in Fig.~2(b),
and peak "B" is within one of those. In contrast to this the peak
"C", with small transmission value, is well described by the model,
as well as the region with transmission close to 1
($V_{QPC2}>-$651~mV) and the one on the right wing of the peak "A"
($-$708~mV$<V_{QPC2}<-$678~mV).

\begin{figure}
\includegraphics[width=85mm]{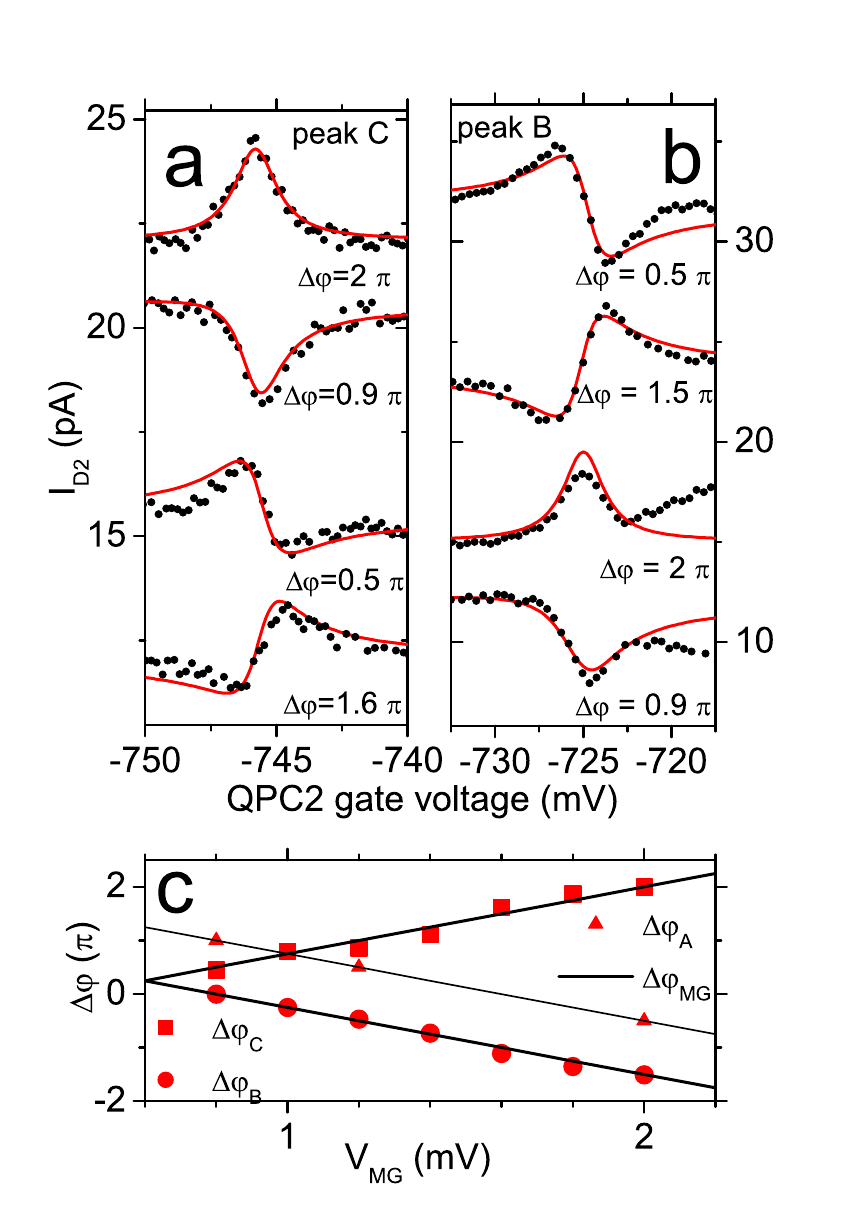}
\caption{(color online) (a)  The resonance "C" (dots), shown in
Fig.~2(c), fitted by model of an interferometer with quantum dot in
one arm (full line) for different voltages of modulation gate
$V_{MG}$=0.4, 0.8, 1.2 and 2.0~mV changing from bottom to top. (b)
The same for the peak "B" in Fig.~2(c). The curves in (a), (b)
shifted for clarity except lowest one. (c) The phase shift between
interferometer arms, found from the fit to the resonances "B" and
"C", plotted as function of $V_{MG}$. The full lines in (c) are
expected slopes resulted from AB period $\Delta V_{MG}$=1.6~mV in
$V_{MG}$. The phase evolution for the resonance "A" was evaluated by
procedure described in the text.}
\end{figure}

Following the noninteracting particle approach from
Ref.~\onlinecite{shuster} we add a quantum dot to the above
mentioned MZI, whose transmission properties are modelled by the
Breit-Wigner formula. We replace transmission and reflection
amplitudes of QPC2 by those of the quantum dot. The dot transmission
amplitude has a phase which must be added to the cosine argument in
Eq.~1. Then the coherent component in MZI transmission is
proportional to $T_{SD2coh}\propto \vert t_{QD}\vert\,\vert
r_{QD}\vert\,\cos[\Delta\varphi+\theta(t_{QD})]$ where $ t_{QD}$,
$r_{QD}$ and $\theta(t_{QD})$ stand for the transmission
(reflection) amplitude of QD and the phase of QD respectively. The
Breit-Wigner formula for the selected state with the energy $E_n$
\begin{equation}\label{BW}t_{QD}=\vert t_{QD}\vert
e^{i\theta}=\frac{(i\Gamma/2)}{(E-E_{n}+i\Gamma/2)}\;,\end{equation}
with $\theta(t_{QD})=\arctan[\frac{2}{\Gamma}(E-E_{n})]$ and
relation $\vert r_{QD}\vert=\sqrt{1-\vert t_{QD}\vert^{2}}$ were
applied for the simulation of experimental curves [in Fig.~4].

This model rather well describes the interference resonances "B" and
"C". The peak "A" was analyzed below only qualitatively, because it
deviates over a wide range from Eq.~\ref{BW}. The resonance "C" was
matched well by calculated curves [Fig.~4(a)] with resonance width
of $\Gamma_{C}$=1.9~mV determined from fit by Lorentzian the peak in
the QPC2 transmission in Fig.~2(a). Therefore the only fitting
parameter for this interference resonance was the phase shift
$\Delta\varphi$. For peak "B" [Fig.~4(b)] an effective transmission
resonance, of smaller amplitude and width [$\Gamma_{B}$=2.5~mV
instead 3.4~mV in Fig.~2(a)], matching the experimental visibility
in the Fig.~2(b), was used.

In addition to the good matching of the experimental curves in
Fig.~4(a,b) to Eq.~\ref{BW}, the validity of our interpretation is
supported by the correlation between the phase set by the modulation
gate voltage and that determined from the best fit to experimental
data in Fig.~4a and b with Eq.~\ref{BW}. The period of AB
oscillations in $V_{MG}$ was found to be 1.6~mV, which corresponds
to phase change of 2$\pi$. In the Fig.~4(c) we plot the phase found from
fitting as function of modulation gate voltage
$\Delta\varphi(V_{MG})$ for the peaks "C"(squares) and "B"
(circles), and from these graphs extract the slope $a$ of
$\Delta\varphi=aV_{MG}$. We find $\Delta\varphi=\pm~2\pi/1.6$~mV,
which is in perfect agreement with the the period $V_{MG}$ extracted
from the interference pattern $I_{D2}(V_{MG})$.

The phase shift $\Delta\varphi$ of peak "A" can only be determined qualitatively as mentioned above. From figures 4(a) and (b) one can see, that a peak in the interference resonance corresponds to $\Delta\varphi=2\pi$, the dip to $\pi$,  combination left peak/right dip to $\pi/2$, and left dip/right peak to $3\pi/2$. Comparing this with the shape of the interference resonance of peak "A" we can deduce an approximate phase shift.
It is interesting to investigate the direction of the phase
evolution in Fig.~4(c) $\Delta\varphi(V_{MG})$ for the neighboring
peaks "A" and "B", and peak "C".

As a result,
peak "A" shows the same direction of phase evolution as "B"
[Fig.~4(c), triangles]. On the other hand, the phase evolution of
peak "C" goes into the opposite direction. This discrepancy may
originate from the variability of two barriers, since the dot is
defined by the single QPC2 gate and the disorder potential. Tuning
the gate potential of QPC2 changes simultaneously the two barrier
heights of the QD and its well depth.

The largest of
the two barriers must be the branching point of the interferometer
path. If the barriers interchange their height, the branching point
will interchange its location as well. In this case the $\pi$ phase
shift from the quantum dot may contribute either to the upper arm or to
the lower one [Fig.~3(b)], and depending on this, gain its different
sign in the paths phase difference.

\section{Conclusions}
In summary we have shown that the frequently
observed transmission resonances in quantum point contacts within an
electronic Mach-Zehnder interferometer stem from inadvertent quantum
dots formed by the disorder potential in high magnetic field and
measured the phase of the transmission amplitude through a quantum
dot. We propose to utilize this effect for the detection of the
state of charge qubits in the vicinity of a Mach-Zehnder
interferometer via their reflection phase. Such a dispersive
read-out may allow more sensitive and less invasive detection than
the currently used quantum point contacts.

\section{Acknowledgements}
We are grateful to K. Kang, K. Kobayashi, S. Ludwig, and T. Hecht
for fruitful discussions. The work was funded by the DFG within the
SFB631 "Solid state quantum information processing" and the BMBF via
project 01BM465 within the program "Nanoquit".

%

%
%
\end{document}